\documentclass[conference]{IEEEtran}
\IEEEoverridecommandlockouts
\usepackage{cite}
\usepackage{amsmath,amssymb,amsfonts}

\usepackage{float}
\usepackage{algpseudocode}
\usepackage{graphicx}
\usepackage{subfigure}
\usepackage{cleveref}
\usepackage[left=1.62cm,right=1.62cm,top=1.9cm]{geometry}

\usepackage{textcomp}
\usepackage{xcolor}
\def\BibTeX{{\rm B\kern-.05em{\sc i\kern-.025em b}\kern-.08em T\kern-.1667em\lower.7ex\hbox{E}\kern-.125emX}}

\begin{document}
\title{ReputeStream: Mitigating Free-Riding through Reputation-Based Multi-Layer P2P Live Streaming}

\author{\IEEEauthorblockN{Rashmi Kushwaha, Rahul Bhattacharyya, Yatindra Nath Singh}
\IEEEauthorblockA{
\textit{Department of Electrical Engineering} \\
\textit{Indian Institute of Technology, Kanpur}\\
Kanpur, India \\
rashmikushwaha221@gmail.com, rahulbhatta0@gmail.com, ynsingh@iitk.ac.in}
}

\maketitle

\begin{abstract}
This paper presents a novel algorithm for peer-to-peer (P2P) live streaming that addresses the limitations of single-layer systems through a multi-layered approach. The proposed solution adapts to diverse user capabilities and bandwidth conditions while tackling common P2P challenges such as free-riding, malicious behavior, churn, and flash crowds. By implementing a reputation-based system, the algorithm promotes fair resource sharing and active participation. The algorithm also incorporates a request-to-join mechanism to effectively manage flash crowds. In addition, a dynamic reputation system improves network efficiency by strategically positioning high-reputation peers closer to video sources or other significant contributors. 
\end{abstract}


\section{Introduction}
Advancements in technology have made internet-connected devices, particularly smartphones, widely accessible. With smartphones becoming integral for both work and leisure, the demand for high-definition video streaming has surged, as professionals increasingly prefer online meetings and conferences. According to Cisco's Visual Networking Index (VNI) \cite{b1}, video is expected to remain the dominant driver of internet data consumption, covering a wide range of formats, from high-definition (HD) and ultra-high-definition (UHD) streaming to live broadcasts and user-generated content. This trend has significantly driven the need for better video streaming solutions.

Clear communication in virtual meetings and webinars requires an adaptable video quality for different speeds and devices on the Internet. HD video enhances engagement, but adaptive streaming is needed to adjust quality in real-time based on bandwidth. As remote work grows, flexible, high-quality video streaming is essential for effective communication and productivity.

The faster Internet has increased the demand for high-quality video, with viewers expecting HD, 4K, and 8K as the norm. As home theaters grow in popularity, streaming services are enhancing video quality by offering more high-resolution content and advanced streaming tech. This ensures a superior viewing experience, helping them attract and retain subscribers in a competitive market.

The rise of video games and virtual reality (VR) has expanded the demand for high-quality video, as modern games and VR experiences require high-definition visuals and fast frame rates for immersion. This demand drives innovation, pushing the video content industry to deliver superior quality to meet the expectations of tech-savvy users.

High-quality video is essential for online education, providing clear visuals for detailed learning and enhancing engagement. Adaptive streaming ensures accessibility across varying internet speeds, allowing students from different regions to benefit from uninterrupted content. HD video also supports interactive tools like screen sharing and virtual labs, enriching the learning experience for all.

Varying internet speeds globally requires adaptive streaming to offer different video qualities. Fast connections enable HD, 4K, and 8K, while slower speeds rely on SD to prevent buffering. This drives innovation, ensuring a good experience for all users, regardless of bandwidth.
Single-layer streaming \cite{b15,b16,b17,b18,b19,b20,single}, does not account for differences in peers' bandwidth and capabilities.

\begin{figure}
    \centering
\includegraphics[width=5cm,height=6cm,keepaspectratio]{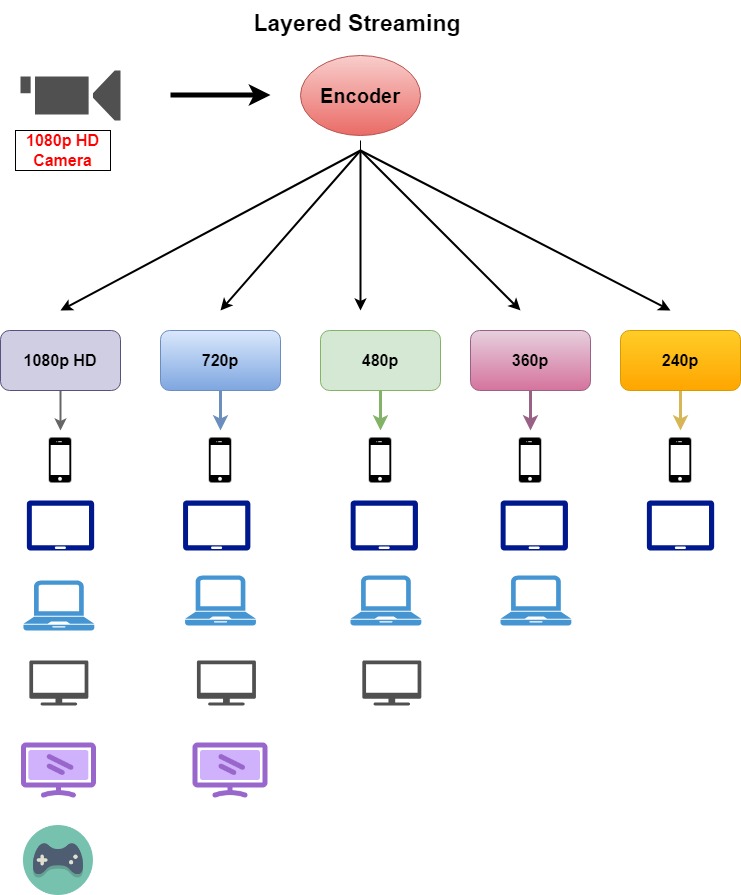}
    \setlength{\belowcaptionskip}{-10pt}
    \caption{Layered encoding}
   \label{fig: layering}
   \end{figure}

To meet these demands, layered streaming in peer-to-peer (P2P) networks is a possible technique. In layered streaming, video is encoded into multiple layers, with essential lower layers and optional higher ones. Peers with higher reputations receive more layers, enjoying better video quality as a reward for contributing more by uploading to others. This method improves streaming performance by enhancing resilience to network issues, supporting scalability, and adapting video quality to network conditions for an optimal viewing experience. Figure \ref{fig: layering} shows layered streaming capable for different types of devices.

Traditional streaming services using a client-server model have significant drawbacks such as scalability, central point of failure, prompting researchers to explore Peer-to-Peer (P2P) streaming architectures \cite{b22,b23,manzato2008incentive,agarwal2007analysis}. In a P2P network, the issue of selfish peers-those who consume resources without contributing-can significantly degrade the overall performance and reliability of the system \cite{b2,b3,b4,b5,b6,b7,b8}. To address this, an incentive mechanism \cite{b11,b12,b14}, is essential to encourage active participation and fair resource sharing among all peers. Various incentive models exist, such as credit-based systems \cite{b13,b24,b25}, and reputation-based systems \cite{b9,b10}. We have implemented reputation as an incentive within our live streaming architecture.

In this system, each peer is assigned a reputation score based on their contributions. Peers with higher reputation scores gain access to a better quality (more layers) video stream and slightly early reception, incentivizing them to contribute more to the network. Conversely, peers with lower scores may receive lower quality streams (with less layers) or face restrictions, motivating them to improve their reputation by contributing more resources. 

The main reason for using layered video \cite{b38,b39,b40,b41,b42,b43}, is to offer stronger incentives for contributing peers in a P2P network. Peers who contribute more resources are rewarded with access to more layers, leading to better video quality. This encourages active participation, as peers seeking higher quality video are motivated to contribute more. By linking contributions directly to viewing quality, the layered video approach promotes a fair and efficient resource distribution, improving both network performance and user satisfaction.

\par In this paper,
\begin{itemize}
\item We have outlined a multi-layer live streaming architecture consisting of five key stages, which uses layered streaming as an incentive to promote active peer participation.
\item We have developed a reputation aggregation formula that incorporates the number of layers, allowing us to monitor peers' reputations based on their contributions.
\item We have established a game-theoretic equilibrium, where the system stabilizes based on peer satisfaction.

\end{itemize}
\par In this paper, Section II covers related work, while Section III introduces our architecture for a reputation-based multi-layer live streaming system. Section IV discusses Nash Equilibrium for ensuring system stability. Lastly, the conclusions are presented.

\section{Related work}
Layered streaming, a technique utilized in P2P networks, significantly improves video delivery by dividing a video stream into multiple layers, each offering incremental quality enhancements. This approach addresses typical challenges of traditional streaming, including bandwidth heterogeneity, network fluctuations, and varying device capabilities. Using scalable video coding (SVC), layered streaming allows a video to be encoded into a base layer and several enhancement layers. The base layer ensures a minimum acceptable quality, while enhancement layers can be added or removed to adapt to network conditions or user capabilities, thereby optimizing the streaming experience.

In a P2P network, the layered streaming approach is particularly beneficial as it allows peers to receive layers based on their contributions. Peers with higher bandwidth and processing power can access and share more layers for better video quality, while those with lower capabilities still participate by sharing fewer layers. Research on this approach addresses bandwidth limitations and ensures diverse peer participation. Studies also focus on managing network dynamics, like churn and flash crowds, with adaptive algorithms that adjust layer distribution. 

Another significant contribution is the use of reputation-based systems in layered streaming. Reward peers who contribute more to the network by uploading or forwarding more layers with access to higher-quality layers. Many authors have investigated the use of reputation-based systems in layered streaming, exploring how these systems incentivize peer contributions and address issues like free-riding.

In \cite{b38}, the authors propose LayerP2P, a system that divides video into multiple quality layers, enabling adaptation to different network conditions and peer capabilities. It uses a chunk-layer scheduling algorithm and supplier-side quality adaptation to optimize video quality and playback. The system is robust to peer churn and scales well with network size. However, while the paper highlights the role of incentive mechanisms in encouraging participation, it does not thoroughly address their effectiveness in preventing free-riding. 

PALS \cite{b39}, is a peer-to-peer streaming system that combines layered video coding with adaptive streaming to improve video playback and quality in dynamic networks. It adjusts the number of layers peers receive based on network conditions and capabilities, ensuring smooth playback. PALS uses a receiver-driven protocol for data delivery and congestion control for fair bandwidth distribution. It enhances bandwidth utilization and video quality compared to non-adaptive methods but does not explicitly address incentive mechanisms for encouraging peer participation.

In \cite{b40}, the authors present a framework aimed at optimizing efficiency, fairness, and incentives in layered P2P streaming. It includes a distributed algorithm to balance video quality and system performance, a chunk scheduling algorithm prioritizing the base layer, and a tit-for-tat-like mechanism rewarding peers who contribute more bandwidth. While the framework promotes fairness and participation, it assumes cooperative peer behavior, which may not hold in real-world scenarios where selfish or malicious actions could undermine its effectiveness.

In \cite{b41}, the authors introduced Layeredcast, a hybrid P2P live video streaming system combining tree-based and mesh-based architectures. The tree structure handles the base layer for low delay, while the mesh manages enhancement layers for flexibility. Intelligent peer selection and layer subscription optimize video quality and efficiency. Layeredcast outperforms single-architecture methods, but its effectiveness depends on peers with sufficient resources, potentially causing unequal quality distribution among participants.

This seminal paper \cite{b42}, introduces an incentive-based system for P2P live streaming using layered video. Peers are rewarded with higher quality layers based on their network contribution, while low contributors receive only the base layer. A distributed algorithm determines layer eligibility based on upload contribution. This approach pioneered the use of quality differentiation as an incentive mechanism in P2P streaming. However, the system's performance is vulnerable to peer churn, as the loss of high-contributing peers can disrupt higher layer availability and degrade the viewing experience for others.

In \cite{b43}, the authors present a resilient P2P streaming system designed to handle peer churn and network disruptions. It employs redundant paths and intelligent peer selection to ensure smooth streaming, even as peers join or leave. The system balances load and minimizes latency for high-quality video delivery, but its focus on resilience may result in suboptimal bandwidth use by prioritizing multiple connections and backup paths over efficiency.

Layered streaming in P2P networks offers a flexible and efficient solution for video delivery in heterogeneous networks. Building on scalable video coding and reputation-based incentives, research in this area continues to evolve, tackling modern network complexities. Future advancements will likely focus on optimizing these systems for diverse and dynamic user environments.

\section{System Architecture}

\subsection{{Five Core Stages of Architectural Framework}}
Our system architecture can be organized into five fundamental stages.
\begin{itemize}
    \item Joining and authentication
    \item Initialization and topology formation
    \item Construction of routing tables
    \item Streaming of media
    \item Reputation calculation and aggregation
\end{itemize}
\vspace{1mm}
\subsubsection{\textbf{Joining and authentication}} After downloading the client software, peers sign up using a username (email ID or mobile number). All peers initially join the base distributed hash table (DHT) layer, a common framework that connects all peers regardless of their specific functionality, supporting various applications such as mailing, messaging, file sharing, and live streaming.

The Authentication Manager (AM) authenticates each user with the authentication server, allowing users to create credentials as digitally signed certificates containing their userID and public key. Users can mutually authenticate each other using SSL protocol with these certificates, digitally sign content, and decrypt content meant for them. The AM ensures that messages are securely transacted without tampering. After authentication, users receive a pair of cryptographic keys (public and private). Each node also has a unique nodeID generated using the public-private key pair to ensure randomness.

When a peer wants to start live streaming, it acts as the source and creates a streamID. A stream descriptor is then generated, containing the streamID, source nodeID, a description of the live content, and the date and time of the stream. The streamID is published with the bootstrap node as the layerID, with an initial Routing Table (RT) containing only the source nodeID. Since the bootstrap node serves as a leaf node in all DHT layers and does not participate in streaming, the initial RT of the streamID is accessible to any interested node. The streamID and descriptor can also be advertised through other channels. Upon creation of the streamID, two corresponding DHT layers are generated and linked.
\begin{itemize}
    \item One for media streaming,
    \item Other for registering the reputation value.
\end{itemize}
     
The source encodes video into multiple dependent layers using a layered coding scheme. Upon stream creation, a blank RT-NT is published with the bootstrap node, which tracks all streams and maintains the RT in the media streaming and reputation layers.

Subscribing peers use the streamID to get RT information from the bootstrap node. After updating their RT, peers create their network for the stream using an RT maintenance algorithm and build an NT based on RTT measurements. Peers in the media streaming DHT are also part of the reputation DHT for efficiency. Stream access is open via streamID, with encryption-based access control.

\subsubsection{\textbf{Initialization and topology formation}} Peers periodically update their Routing Table (RT) and Neighbor Table (NT) by exchanging entries with listed nodes to establish a dedicated streaming network. When streaming begins, the source sends beacons to peers in its RT and NT. These beacons, containing the source's nodeID, beacon sequence number, and expiry timer, help maintain connectivity and identify potential parent nodes. Each node selects the peer from which it first receives a beacon with a new sequence number as its parent.

Subscriber nodes relay beacons to all known nodes, excluding those from previously received copies. Duplicates are identified and discarded based on their sequence numbers. When peers receive the first beacon with a new sequence number, they record the sender as a parent node in their Broadcast Routing Table (BRT). If they wish to join the stream, they send a request to this parent. The parent either accepts the request or, if unable to accommodate it, provides alternative node IDs (its child nodes). Accepted peers are added as children, forming an overlaid tree network. This tree structure facilitates the distribution of media packets from the source to all receiving peers, with each peer maintaining five distinct tables to manage the network topology.

\subsubsection{\textbf{Construction of routing tables}} Once a topology is established for each streamID, each peer maintains five tables. These tables help peers share their media feed, remain connected during disruptions, and provide backup peers in case of churn events.

\begin{enumerate}
\item \textbf{Routing Table (RT)}- This table is created for each streamID (layerID) and stores entries of participating nodes, with a maximum of 120 nodeIDs as per our DHT algorithm. Initially populated by the bootstrap node, it is later maintained using the RT maintenance algorithm. Our architecture incorporates routing tables (RT) for both media streaming and the corresponding reputation layer.

\item \textbf{Neighbour Table (NT)}- This table is initially created using the routing table (RT), recording up to 16 nodeIDs closest to the node in terms of RTT (Round Trip Time). In our architecture, the neighbor table (NT) is maintained exclusively for media streaming and is periodically updated by exchanging RT-NT entries, keeping only the lowest RTT entries and discarding the rest.

\item \textbf{Broadcast Routing Table (BRT)}- This table tracks nodeIDs from which beacons are received and maintains entries for the top three nodes based on beacon consistency. This approach ensures system resilience by providing redundancy if a node fails. Once populated, a join request is sent to the primary parent node, with a fallback to the next listed node if no response is received within a specified timeframe.

\item \textbf{Overlaid Multicast Table (OMT)}- This table tracks the parent peer from whom the media feed is received and the child peers to whom the feed is sent. While there is only one parent peer, multiple child nodes can receive the stream. Join requests are sent sequentially to parent nodes based on the BRT, with the OMT recording the parent that accepts the request. If no response is received before the timer expires, the request is resent. Child nodes that fail to send a join request within the allotted time are removed, and the timer resets upon receiving a join request.

\item \textbf{Backup Parents Table (BPT)}- This table stores information about grandparents and siblings to maintain network connectivity during disruptions. When a peer accepts a join request, it provides details about its parent (grandparent); if it rejects the request, it shares information about its children (potential siblings). The requesting peer stores this data in the BPT, which facilitates efficient recovery during churn or network issues.
\end{enumerate} 

\subsubsection{\textbf{Streaming of media}}
The source node encodes and sends all layers of the video, but the number of layers that a peer receives depends on its reputation and position within the network topology. Since any upper layers cannot be decoded without all the lower layers, peers with higher reputation and better network positioning receive more layers, resulting in better video quality.

\begin{figure}
    \centering
    \includegraphics[width=08cm]{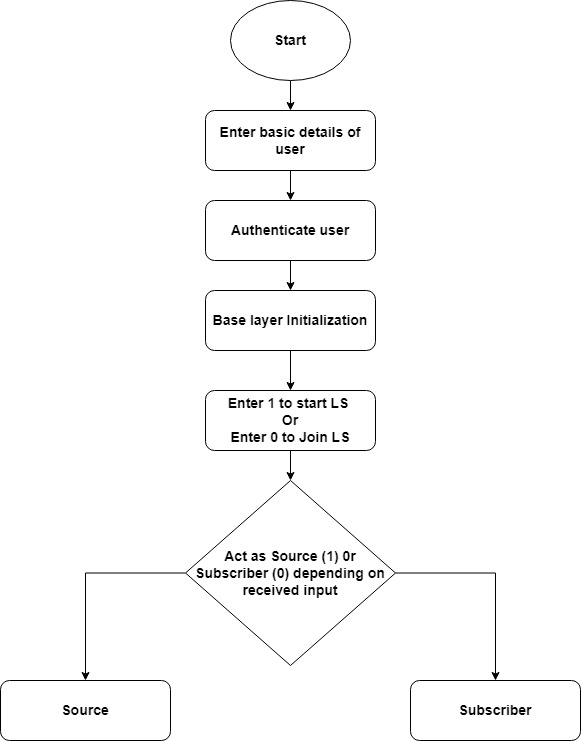}
    \setlength{\belowcaptionskip}{-8pt}
    \caption{Flowchart of live streaming working}
   \label{fig: flowchart1}
   \end{figure}

\begin{figure}
    \centering
    \includegraphics[width=08cm]{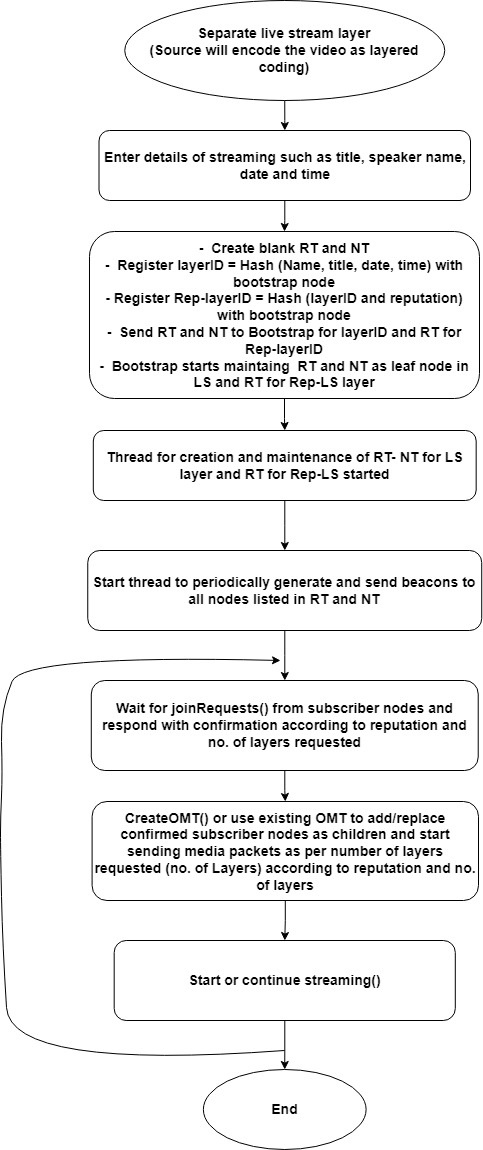}
    \setlength{\belowcaptionskip}{-8pt}
    \caption{Streaming Setup and Management for Source}
   \label{fig: flowchart2}
   \end{figure}

\begin{figure}
    \centering
    \includegraphics[width=08cm]{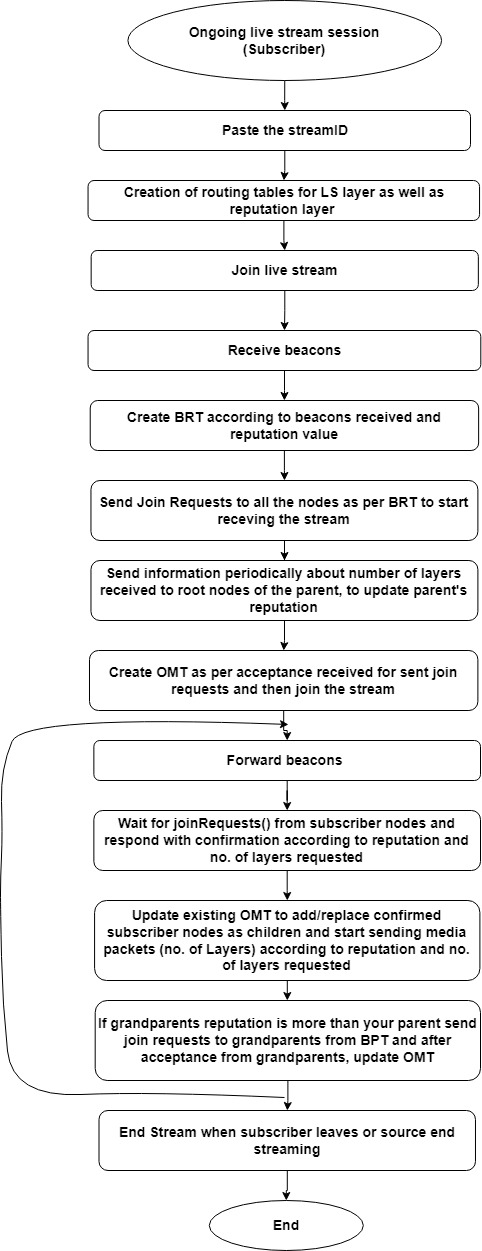}
    \setlength{\belowcaptionskip}{-10pt}
    \caption{Streaming Setup and Management for Subscriber/forwarder}
   \label{fig: flowchart3}
   \end{figure}

A peer in the network can function either as a source or a subscriber. Figures \ref{fig: flowchart1}, \ref{fig: flowchart2}, and \ref{fig: flowchart3} show a flow chart for the setup and management of the medium streaming for the source and subscriber.

When the user opens the application, they can enter their details such as name, email ID, alias, and password, which are then sent to the authentication server for verification. Once verified, the base layer is initialized, connecting all active peers in the network. The user is then given the option to either start a live stream or join an existing one. Depending on the user’s choice, the application follows one of two distinct pathways, with the user’s node either functioning as a source node or a subscriber node, each leading to different operational processes.

When a source user initiates live video streaming, a distinct live stream layer is created, along with a corresponding reputation layer. The source node encodes the video into layered video streams. The layerID for the streaming layer is generated by the source node using descriptive parameters such as the source nodeID, date, and starting time, which are hashed together (e.g., layerID = hash(name, title, date, and time)). For the reputation layerID, a hash of the layerID and reputation is used. The source node then creates blank Routing Tables (RT) and Neighbor Tables (NT), and registers the live stream layerID and the corresponding reputation layerID with the bootstrap node, and sends RT and NT to the bootstrap node before starting the live stream session. The bootstrap node, acting as a leaf node, begins maintaining RT and NT for the live stream and RT for the corresponding reputation layer. A thread is initiated for the continuous creation and maintenance of these tables.

Any new node that wants to join the live stream layer will use the streaming layerID and reputation layerID to retrieve the routing tables from the bootstrap node. Afterward, the node integrates itself into the live stream layer and the corresponding reputation layer using the routing table management process.

The source node periodically generates beacons containing a sequence number and an expiry timer, which are sent to all nodes listed in the RT and NT of the live stream layer. Each node forwards the beacon if it is the first with an unseen sequence number and records this sequence number with the timer value. If a new sequence number is not received before the timer expiry, the entry is removed from the broadcast routing table. Beacons are forwarded only to the nodes in the same layer from whom the node have not yet received a copy. Each node records the node from which the first beacon was received as the next hop towards the source node.

The source node then waits for join requests from other nodes that wish to receive the stream. Upon receiving a join request, a node checks its capacity to forward the stream. If it has the capacity, the node adds the requester to its child table and begins transmitting media packets. If the capacity is exceeded, the node compares the no. of layers requested and the reputation of the requesting node with its existing children. If the requesting node has a higher reputation*no. of layers requested than the lowest reputation child, the requesting node replaces the existing child. The replaced child is informed about other potential parent nodes from which it can receive the stream. The stream is then copied to all the child nodes. After joining, the requesting node submits a reputation update message periodically along number of layers being received to the root of its current parent node in the reputation layer.

Media streaming continues based on the entries in these tables until the source decides to stop streaming.

When a user wishes to join an existing live stream session as a subscriber, the streamID for the session should have been shared through one of the various means like message group, direct message, websites etc.. The user is prompted to input the streamID in the beginning. Once the streamID is obtained, the user retrieves the routing tables for both the live stream (LS) layer and the reputation layer for the stream from the bootstrap node. Initially, the user requests a default reputation value from the root nodes in the reputation layer. The reputation will increase or decrease depending on how as the user contributes the resources to the network by providing services to other nodes.

Next, the subscriber node exchanges routing tables with other nodes in the layer to populate and optimize its RT and NT tables. It then begins receiving beacons from source nodes through its parent nodes. Based on these beacons, a beacon routing table (BRT) is created, which is optimized by maintaining entries for only the top three nodes from which beacons are received first. This increases the likelihood of the subscriber receiving the media stream from these BRT entries.

The subscriber node sends join requests to all three potential parent nodes listed in the BRT and waits for responses. Once a join request is accepted by any of the nodes, an overlaid multicast table (OMT) is established, with the accepting node becoming the subscriber’s parent node. The subscriber node then starts receiving the media stream from this parent node. Additionally, the subscriber forwards received beacons, replacing the parent node’s ID with its own, thus acting as a beacon forwarder and propagating the media stream to other nodes.

Subscriber nodes also handle join requests, update their overlaid multicast tables, and transmit the received stream to their children nodes. They send join requests to grandparent nodes listed in the BRT, and upon acceptance, they update their OMT accordingly, forming a tree-like structure that helps distribute the load from the source node. The stream for the subscriber concludes either when the subscriber node leaves or when the source node stops streaming. After the stream ends, all nodes in the live stream and reputation layers remove the corresponding RT and NT, stop periodic updates, and exit the layer.

\subsubsection{\textbf{Reputation calculation and aggregation}} A separate layer has been established to store and update reputation, linked directly to the corresponding streamID. New client peers are initially assigned a base reputation value upon joining. When a peer receives a request from another peer, it first evaluates the requester’s reputation value, which is retrieved from the reputation DHT layer and also no. of layers requested by him. The peer then decides whether to accept or reject the request based on this factor i.e. reputation value*no. of layers.

\textbf{Reputation aggregation formula} can be represented as;

\begin{equation}
    R(t)=\frac{R(t-\tau) e^{-\alpha \tau}+R_rL}{1+R_rL},
\end{equation}
\hspace{5mm} where, 

$R_r=$ \textit{Reputation of receiving peers who is sending reputation value},

$L=$ \textit{No. of layers},

$R(t-\tau)=$ \textit{Last estimated value of reputation at time \lq$(t-\tau)$\rq},

$\alpha=$ \textit{Decay factor (rate at which reputation value will decay or forgotten)}. 

\textit{The value of $R(t)\in$ [0,1]}.

As peers transmit more layers, their reputation increases, granting access to higher video quality. Peers can further boost their reputation by sending video to higher-reputation requesters. In this dynamic network, nodes prioritize connecting to high-reputation peers. 

\textit{\textbf{Argument for the value of $\alpha$}: In live streaming, which is an ongoing process with a finite duration, rapidly reducing the reputation of free riders is essential for maximizing system performance. To achieve this, a larger value of alpha is used to accelerate reputation decay. This ensures that free riders are quickly identified and relegated to the bottom of the tree, thereby improving overall performance during the streaming.}

The reputation DHT layer maintains three identical copies of reputation data distributed across different peers to prevent any single node from tampering with the reputation values. Consequently, each peer typically stores the reputation data of three other peers on average.

Once streaming starts, each node creates the necessary tables and begins supplying media feeds to its child peers. These child peers then send reputation updates to the root nodes of their feeding peers in the reputation DHT. The root nodes update the reputation of the feeder peers based on the received information. 

Peers use these updated reputation values to decide whether to accept or reject requests, favoring those with higher reputations. Additionally, if a peer is streaming to multiple others, it can check the reputation DHT to confirm that all peers are submitting accurate reputation updates. This process ensures accountability and correctness in reputation reporting. 

\begin{figure}
    \centering
    \includegraphics[width=08cm]{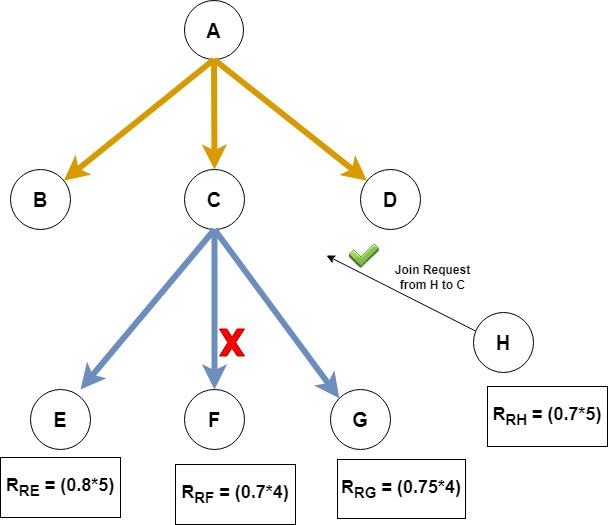}
    \setlength{\belowcaptionskip}{-10pt}
    \caption{Topology}  
   \label{fig: T3}
\end{figure}

During streaming, if a higher-reputation node requests a connection, it replaces a lower-reputation child. The parent node then provides the displaced child with a list of its former siblings, offering potential new parent connections. This process continually reorganizes the network, ensuring high-reputation nodes occupy key positions, promoting efficient resource distribution, rewarding contributions, and encouraging active participation. Figure \ref{fig: T3} shows this scenario.

Suppose that $H$ has a higher factor i.e (reputation*no. of layers requested) than $F$, so $C$ will replace $F$ with $H$, but before replacing, $C$ will tell $F$ about his children's node ($E$, $G$ and $H$).

Suppose that a sender peer {\lq{A}\rq} sends media feed to some of the peers {\lq{B}\rq}, {\lq{C}\rq}, {\lq{D}\rq} and {\lq{E}\rq}. Number of layers between a parent {\lq{P}\rq} and child {\lq{C}\rq} is minimum of what {\lq{P}\rq} is receiving and what it can be transmit to {\lq{C}\rq}, so that $C's$ playback buffer is neither empty nor full. This is to be dynamically adjusted by node {\lq{C}\rq}. 

Let's suppose we have 5 layers of video. In the example, let the value of L lies in between 2 and 5 i.e. $1<L<=5$. 

Let reputation value of {\lq{B}\rq}, {\lq{C}\rq}, {\lq{D}\rq} and {\lq{E}\rq} be $R_{rB} = 0.9, R_{rC} = 0.7, R_{rD} = 0.8, R_{rE} = 0.85$ respectively. Also layers fed to {\lq{B}\rq}, {\lq{C}\rq}, {\lq{D}\rq} and {\lq{E}\rq} be 5, 3, 4 and 4 respectively.

So, we have $$M=\frac{\sum_{i=1}^KR_{r i}L_i}{K}$$

So, $
\begin{aligned}
& M=\frac{(0.9 \times5)+(0.7 \times3)+(0.8 \times4)+(0.85 \times4)}{4}
\end{aligned}
$ 

So, we have reputation updated at time $t$, from earlier estimate at $(t-\tau)$, as 
$$R(t)=\frac{R(t-\tau) e^{-\alpha \tau}+M}{1+M}.$$

\textit{Argument: In this context, it can be concluded that peers streaming to a greater number of high-reputation peers, with the more number of layers, are positioned closer to the source within the network topology. As a result, these peers are likely to receive higher quality video streams}.

\section{{Stability of the system}}
\hspace{5mm}We have defined the stability of the system using the Nash equilibrium in game theory. Game theory is a mathematical framework for analyzing strategic interactions among rational decision makers, where outcomes depend on the choices of all participants. 
There are two types of peers in the system:
\begin{itemize}
    \item Altruistic Peers (Good peers) 
    \item Free riders and Malicious peers (Bad peers)
\end{itemize}

\textit{Assumption: Majority of peers are good}.

\textbf{For Good Peers-} Altruistic peers report accurate reputation values and serve according to their capacity without any strategic manipulation. They adhere to the system's rules and contribute effectively, ensuring the system's proper functioning.

\textbf{For Bad Peers-} 
\begin{enumerate}
\item Free Riders: When interacting with altruistic peers, free riders have their true reputation reported accurately, making them easy to identify based on their reputation. When interacting with other bad peers, they might either report their true reputation or provide misleading information. However, our system uses an aggregation formula that prioritizes the majority of accurate reports from altruistic peers, minimizing the impact of dishonest information from bad peers. Malicious groups do not influence this process significantly.

\textit{Malicious collective will not play any role.}

\item Malicious Peers: These peers send false reputation values when interacting with altruistic peers. Despite this, the system’s aggregation mechanism relies on the majority of honest reports, which helps counteract the effects of malicious behavior. Additionally, our system allows peers to verify the reputation values received from others.
\end{enumerate}  
    
\begin{figure}
    \centering
    \includegraphics[width=04cm]{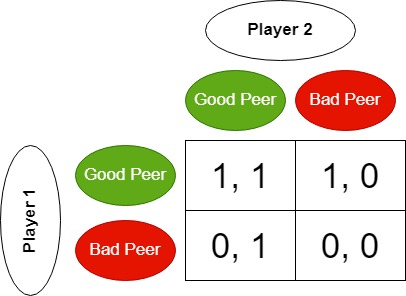}
    \setlength{\belowcaptionskip}{-10pt}
    \caption{Payoff Matrix}
    \label{fig: Payoff Matrix}
   \end{figure}

Now, according to the above scenarios, if we draw a payoff matrix then, we have figure \ref{fig: Payoff Matrix}.

\textbf{1}- Their reputation value contributes to the aggregation process, meaning that the reputation they send is taken into account based on the majority consensus.

\textbf{0}- Their reputation value is ignored in the aggregation process, meaning that the reputation they send does not impact the overall aggregation.

\textbf{What equilibrium can be achieved in the system?} At equilibrium, a peer will not alter its strategy if it results in a higher payoff. Deviating from this equilibrium would result in a disadvantage.

\textbf{Altruistic Peers}- They benefit by actively serving, which keeps their reputation high. Their accurate reporting contributes positively to the reputation aggregation process.

\textbf{Free riders}- They incur losses due to their lack of service, leading to a declining reputation and potential expulsion from the system over time.

\textbf{Malicious peers}- They face disadvantages as their false reputation values do not affect the reputation aggregation, due to the predominance of honest peer contributions. Their malicious efforts are thus largely ineffective.

   \begin{figure}
    \centering
    \includegraphics[width=04cm]{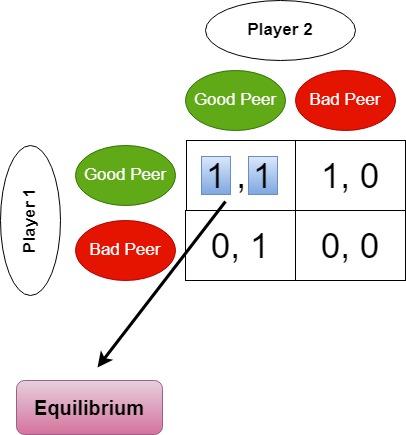}
    \setlength{\belowcaptionskip}{-10pt}
    \caption{Nash equilibrium}
   \label{fig: Nash equilibrium}
   \end{figure}

Figure \ref{fig: Nash equilibrium}, shows the Nash equilibrium. From the above observations, we see that, at equilibrium, free riders and malicious peers will be excluded from the system, while good peers will receive higher payoffs for their contributions. This leads to an increase in the reputation of altruistic peers, potentially reaching the maximum value of 1, and ensures their demands are met. With satisfied altruistic peers, the system’s performance will improve, resulting in better streaming quality and reduced delays, which are critical for live streaming. Ultimately, the system will stabilize, reflecting enhanced peer satisfaction and overall efficiency.

In layered streaming, peer satisfaction is tied to the number of video layers received, which depends on their contributions. This system offers differentiated quality of service, where higher contributions lead to better video quality. The equilibrium ensures not only network stability, but also encourages active contributions, driving higher satisfaction, and creating a balanced, efficient network.

\section{Conclusion}
Our algorithm overcomes the limitations of P2P single-layer streaming by utilizing a multi-layered approach that adjusts to varying user capabilities and bandwidth conditions. It addresses issues such as free-riding, malicious behavior, churn, and flash crowds through a reputation-based system, promoting fair resource sharing and active participation. Unlike single-layer systems with uniform quality, our algorithm offers multiple video quality layers, rewarding peers with higher contributions with superior streams. This layered strategy not only motivates greater peer contribution but also accommodates users with different capabilities, improving fairness and performance. For handling flash crowds, the algorithm incorporates a request-to-join mechanism. The dynamic reputation system enhances network efficiency by positioning high-reputation peers closer to video sources or other high-contributing nodes, thereby improving service quality. Overall, our solution provides a scalable, adaptable streaming experience that aligns video quality with peer engagement and resource sharing, creating a more effective and collaborative P2P streaming ecosystem.

\bibliographystyle{unsrt}
\bibliography{mybibfile}

\end{document}